\documentclass[prb,twocolumn,twocolumngrid,floatfix]{revtex4}
\usepackage{graphicx}
\usepackage{amsfonts}
\usepackage{amsmath}
\usepackage{bm}
\usepackage{alltt}
\usepackage{fancyhdr}

\begin{document}

\title{Anomalous Workfunction Anisotropy in Ternary Acetylides}

\author{Joseph Z. Terdik$^{1}$, K\'aroly N\'emeth$^{1,{\ast}}$, Katherine C. Harkay$^{1}$, 
Jeffrey H. Terry, Jr.$^{1,2}$,\\ Linda Spentzouris$^{1,2}$, Daniel Vel{\'a}zquez$^{1,2}$,
Richard Rosenberg$^{1}$ and George Srajer$^{1}$}

\affiliation{$^1$Advanced Photon Source, Argonne National Laboratory,
Argonne, Illinois 60439, USA}

\affiliation{$^2$Illinois Institute of Technology, Chicago, IL 60616 USA}

\date{\today}

\begin{abstract}
Anomalous anisotropy of workfunction values in 
ternary alkali metal transition metal acetylides is reported. 
Workfunction values of 
some characteristic
surfaces in these emerging semiconducting materials 
may differ by more than $\approx$ 2 eV as predicted by Density Functional Theory calculations. 
This large anisotropy is a consequence of the relative orientation of rod-like [MC$_{2}$]$_{\infty}$ 
negatively charged polymeric subunits and the surfaces, with M being a transition metal or metalloid 
element and C$_{2}$ refers to the acetylide ion C$_{2}^{2-}$, 
with the rods embedded into an alkali cation
matrix. It is shown that the conversion of the seasoned Cs$_{2}$Te photo-emissive
material to ternary acetylide Cs$_{2}$TeC$_{2}$ results in substantial reduction
of its $\approx$ 3 eV workfunction down to 1.71-2.44 eV on the Cs$_{2}$TeC$_{2}$(010) 
surface while its high quantum yield is preserved. 
Similar low workfunction values are predicted for other ternary acetylides as well,
allowing for a broad range of 
applications from improved electron- and light-sources to 
solar cells, field emission displays, 
detectors and scanners. 
\end{abstract}


\maketitle
\section{Introduction}
For many photo-physical applications photoemissive materials are sought after that
can turn a high fraction of the incident photons into emitted electrons, i.e. materials
that have a high quantum-yield. Often, the quantum-yield of these materials depends
heavily on the wavelength of incident photons. For many applications, ranging from 
electron-guns for synchrotrons and free-electron lasers to night vision devices, 
high quantum-yield photoemission using visible or infrared irradiation is desirable. In 
electron-guns of synchrotrons and free-electron lasers, emission in the visible range is
advantageous for the improved control of the shape of the emitted electron bunch that is critical
for time-resolved applications. In night-vision devices a very low flux of infrared
photons has to be turned into emitted electrons with a high yield in order to obtain an image as
sharp as possible. Therefore there is a quest for new and improved materials with
optimized quantum-yield and low-workfunction \cite{DHDowell10}.

Cs$_{2}$Te has been known since the 1950-s for its high quantum-yield \cite{ETaft53} using
ultraviolet illumination with photon-energies above $\approx$ 3.0 eV and has been used
for many decades as a primary high-yield photocathode. 
Besides not being
photoemissive in the visible region, its other main drawback is that its surface gets
oxidized in practical vacuum whereby 
its quantum efficiency substantially reduces \cite{AdiBona96}.
Despite this disadvantage, Cs$_{2}$Te still has 20-30 times longer operational
lifetime than competing multi-alkali antimonide photocathodes, such as K$_{2}$CsSb
and (Cs)Na$_{3}$KSb, especially when operated in
radio-frequency accelerating cavities \cite{AdiBona96}.

In the process of attempting to design modifications of Cs$_{2}$Te with lowered
workfunction and preserved high quantum-yield we have considered the effects of small gas
molecules on Cs$_{2}$Te surfaces. 
%
\begin{figure}[b!]
\resizebox*{3.4in}{!}{\includegraphics{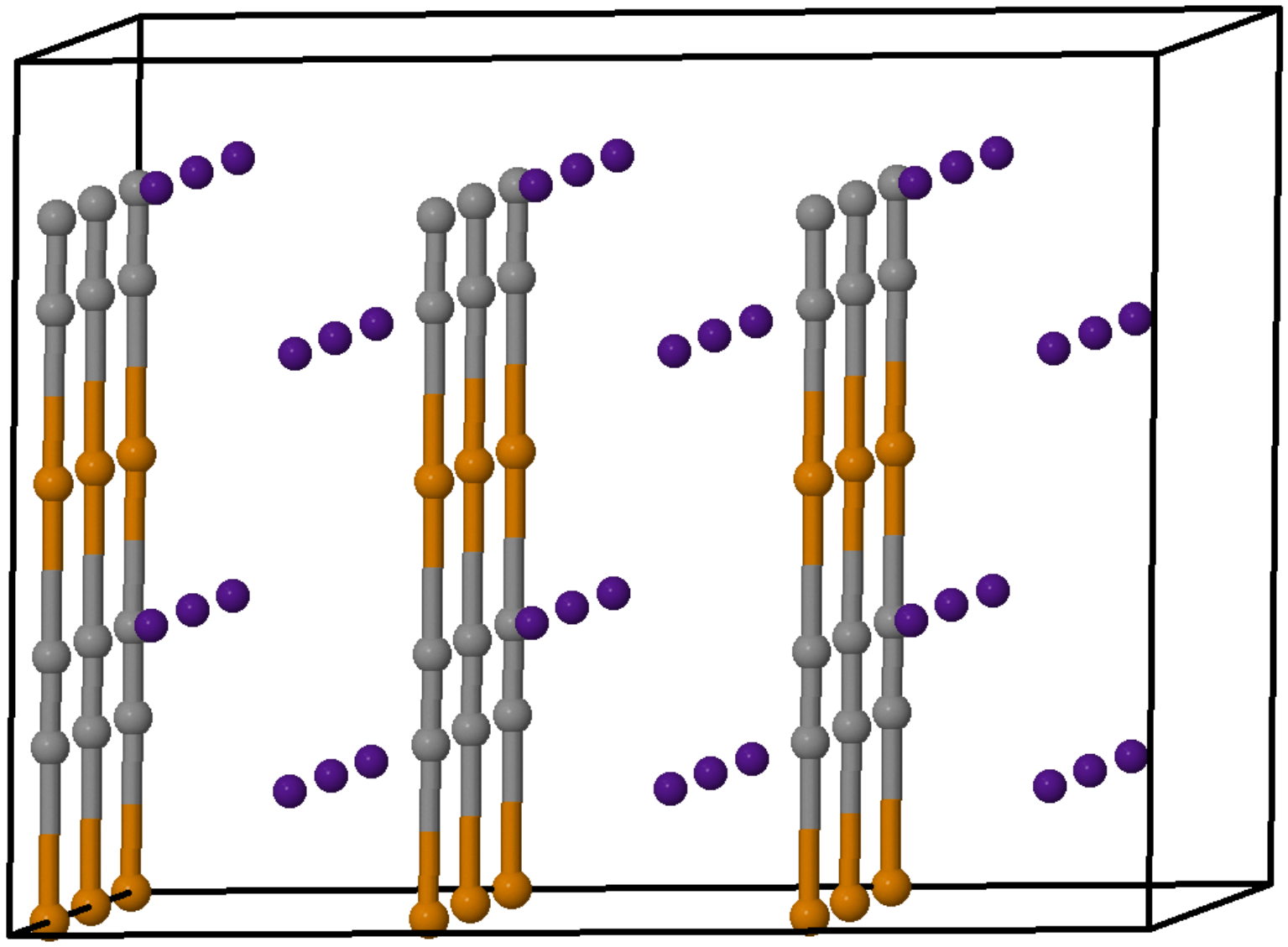}}
\caption{
A side-view of the 3x3x2 supercell of the hexagonal unit cell of Cs$_{2}$TeC$_{2}$.
Bronze spheres denote Te, grey ones C, dark-purple ones Cs.
Notice the [TeC$_{2}$]$_{\infty}$ rods.
}
\label{Cs2TeC2unitcell}
\end{figure}
Such effects have been studied
by di Bona {\it et al.} \cite{AdiBona96}, using small  
gas molecules occurring in vacuum, such as O$_{2}$, N$_{2}$, CO$_{2}$, CO and CH$_{4}$.
It occurred to us that
the effect of another small gas molecule, acetylene (C$_{2}$H$_{2}$) has not been
considered yet, despite the potentially interesting 
reactions between C$_{2}$H$_{2}$ and Cs$_{2}$Te.   
C$_{2}$H$_{2}$ is widely used for welding 
(not in accelerators though)
and it might occur 
in accelerator-vacuums as well, in trace amounts. 
It is an acidic compound and prefers to
decompose to acetylide anion C$_{2}^{2-}$ and to 2H$^{+}$ in the presence
of a base. 
%
\begin{figure}[t!]
\resizebox*{3.4in}{!}{\includegraphics{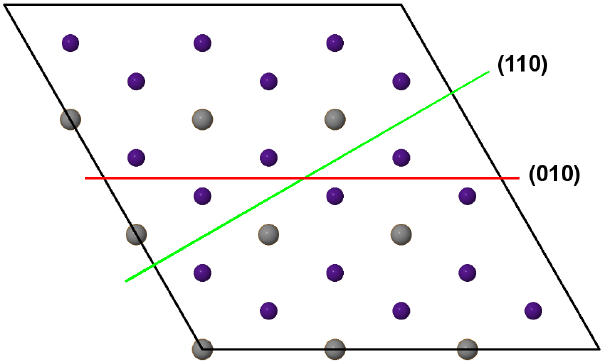}}
\caption{
A top-down view of a 3x3x2 supercell of the hexagonal unit cell of Cs$_{2}$TeC$_{2}$.
Color codes are identical with those in Fig. \ref{Cs2TeC2unitcell}.
The [TeC$_{2}$]$_{\infty}$ rod-like substructures are running perpendicularly to the
plane viewed. The red line indicates the energetically preferred cleavage-plane for the (010)
surface running between two layers of Cs atoms, parallel with the rods,
while the green line refers to the preferred cleavage-plane for the (110) surface that
involves [TeC$_{2}$]$_{\infty}$ rods directly exposed on the surface. Note that the (010)
and (100) planes are identical.
}
\label{Cs2TeC2surfaces}
\end{figure}
%
Based on the acidic character of C$_{2}$H$_{2}$, one
might investigate the working hypothesis that the reaction of 
\begin{equation} \label{Cs2TeC2plusC2H2}
{\rm Cs}_{2}{\rm Te(cr)} + {\rm C}_{2}{\rm H}_{2}{\rm (g)} \rightarrow 
{\rm Cs}_{2}{\rm TeC}_{2}{\rm (cr)} + {\rm H}_{2}{\rm (g)} 
\end{equation}
would produce a ternary acetylide Cs$_{2}$TeC$_{2}$ whereby the oxidation number
of Te would change from -2 to 0 and that of H from +1 to 0, 
with (cr) denoting crystal and (g) gas phase. 
Interestingly, the class of ternary (i.e. three-component) 
acetylides indeed exists, involving already
synthesized members with the general formula of A$_{2}$MC$_{2}$ with A$\in$[Na,K,Rb,Cs]
and M$\in$[Pd,Pt], and the oxidation number of the metal M in them is zero 
\cite{URuschewitz01A,HBilletter10}. 
All existing compounds of the A$_{2}$MC$_{2}$ formula have a hexagonal
unit cell with rod-like [MC$_{2}$]$_{\infty}$ substructures running parallel with the main 
crystallographic axis, and very similar distribution of alkali atoms around the
[MC$_{2}$]$_{\infty}$ rods, just as indicated in Figs. \ref{Cs2TeC2unitcell} and
\ref{Cs2TeC2surfaces}. All known A$_{2}$MC$_{2}$ materials are colored semiconductors 
with 2.1-2.8 eV direct band-gaps \cite{HBilletter10}. The other class of ternary acetylides
with synthesized members contains only a single alkali atom and has the formula of
AMC$_{2}$  \cite{WKockelmann99} with the [MC$_{2}$]$_{\infty}$ rods adopting 3 different kinds
of rod-packings \cite{URuschewitz06}.

\section{Methodology}
\begin{figure}[h]
\resizebox*{3.4in}{!}{\includegraphics{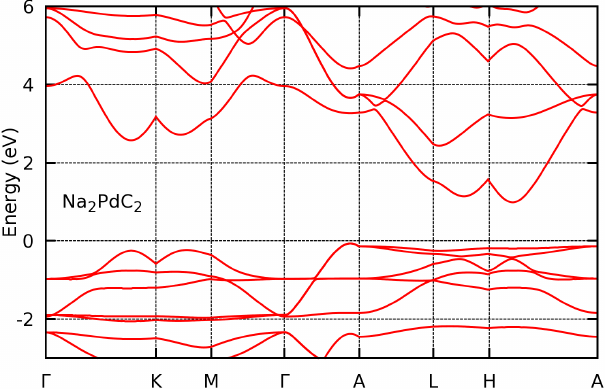}}
\resizebox*{3.4in}{!}{\includegraphics{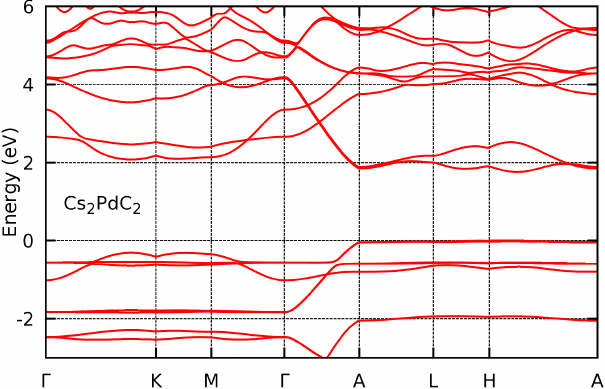}}
\caption{
Bandstructures of Na$_{2}$PdC$_{2}$ and
Cs$_{2}$PdC$_{2}$ using the PBE \cite{PBE} exchange-correlation functional.
The k-space was $14{\times}14{\times}14$ large. The Fermi energy is at 0 eV.
Flat bands are characteristic for ternary acetylides.
}
\label{bands-Pd}
\end{figure}

\begin{figure}[t!]
\resizebox*{3.4in}{!}{\includegraphics{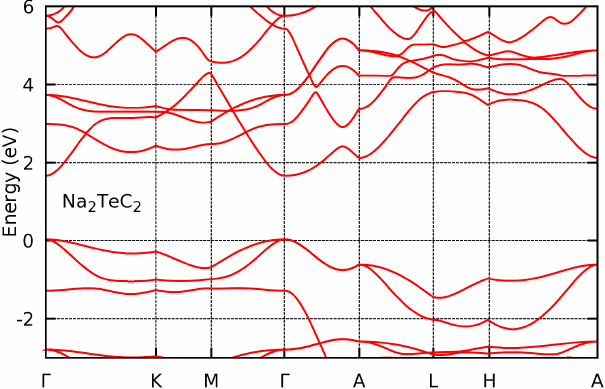}}
\resizebox*{3.4in}{!}{\includegraphics{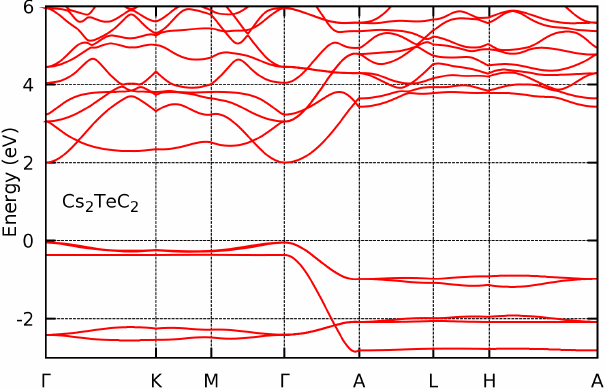}}
\resizebox*{3.4in}{!}{\includegraphics{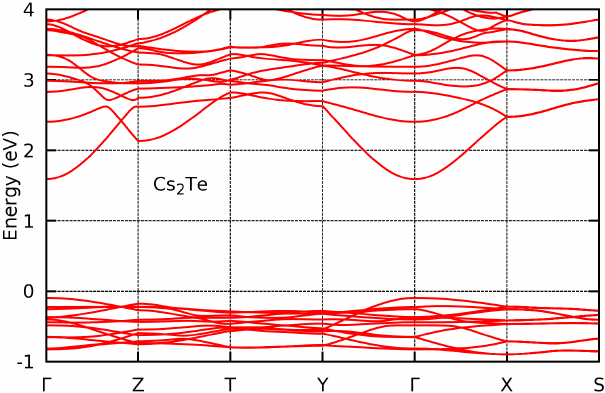}}
\caption{
Bandstructures of Na$_{2}$TeC$_{2}$,   
Cs$_{2}$TeC$_{2}$ and Cs$_{2}$Te, using the PBE \cite{PBE} exchange-correlation functional.
The k-space was $14{\times}14{\times}14$ large. The Fermi energy is at 0 eV.
}
\label{bands-Te}
\end{figure}

Adopting the structure of the unit cell of Na$_{2}$PdC$_{2}$ and substituting Na with Cs and Pd
with Te we have carried out a full crystal structure (lattice parameters and atomic fractional
coordinates) optimization using Density Functional Theory
(DFT), without any symmetry and point group constraints on the translational unit cell. 
We have used the PWSCF-code \cite{QE}, plane-wave representation of wave-functions
with $80$ Rydbergs wavefunction-cutoff, the PBE exchange-correlation functional \cite{PBE} 
in conjunction with norm-conserving pseudopotentials for Cs, Na and Te and ultrasoft
ones for the other elements as available in the PWSCF distribution. 
The k-space grids were at least 6$\times$6$\times$6 large for optimizations, 
the residual forces on fractional
coordinates were less than $4{\times}10^{-4}$ Ry/au, residual pressure on the unit cell
less than 1 kbar.
For validation of the DFT-based methodology, we have calculated known structural parameters and
workfunctions of compounds with similar composition, achieving good agreement
between computed and experimental values as indicated in Tables \ref{StructuralParams},
\ref{Bondlengths} and \ref{ValidationWorkfunctions}.
Note that in some cases, like Na$_{2}$PdC$_{2}$ and Cs$_{2}$PdC$_{2}$ 
the difference between calculated and experimental
 a and b lattice parameters (rod distances) was about 3-3.5\%, significantly larger than that 
for the c lattice parameter ($<$1\%), which we have accepted on the basis that the 
inter-rod interactions are more difficult to accurately predict, similarly 
to general intermolecular interactions.
The workfunction calculations were based on slabs of at least 30 {\AA} width separated by vacuum
layers up to 120 {\AA} following the methodology of Ref. \onlinecite{CJFall99}.
For additional validation of the use of the PBE functional here, we have compared the direct
bandgaps of Na$_{2}$PdC$_{2}$ and Cs$_{2}$PdC$_{2}$ to experimental data. 
Experimental direct bandgaps of
Na$_{2}$PdC$_{2}$, K$_{2}$PdC$_{2}$ and Rb$_{2}$PdC$_{2}$ are at 2.09, 2.55 and 2.77 eV,
that of Cs$_{2}$PdC$_{2}$ is estimated to be slightly greater than that of Rb$_{2}$PdC$_{2}$
\cite{HBilletter10}. 
Our PBE calculations predict the lowest energy direct transitions between 1.2-1.8 eV for
Na$_{2}$PdC$_{2}$ (near the H point) and 1.7-2.6 eV for Cs$_{2}$PdC$_{2}$ (near the
H and K points), as shown in Fig. \ref{bands-Pd}. The overall characteristics of
these bands is similar to those calculated previously for ternary acetylides, e.g.
in Refs. \onlinecite{HBilletter10},   
\onlinecite{URuschewitz06} and \onlinecite{SHemmersbach01}.
Band gaps of bulk Cs$_{2}$Te, Cs$_{2}$TeC$_{2}$ and Na$_{2}$TeC$_{2}$ have been predicted to be
between 1.8-2.0 eV, using the PBE functional (Fig. \ref{bands-Te}).

\begin{figure}[t!]
\resizebox*{3.4in}{!}{\includegraphics{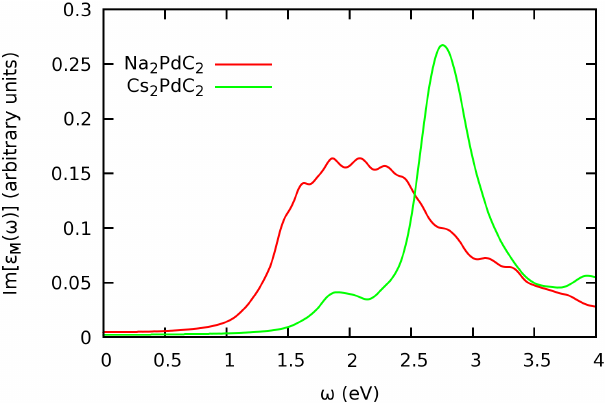}}
\caption{
Optical absorption spectra of bulk Na$_{2}$PdC$_{2}$ and
Cs$_{2}$PdC$_{2}$ from the imaginary part of the macroscopic
dielectric function $\varepsilon_{M}(\omega)$.
Only the G=0 planewaves were used to calculate intensities in the RPA
approximation. The polarization vector of the light is along the main
crystallographic axis (along the [PdC$_{2}$]$_{\infty}$ chains).
}
\label{OptAbs-Pd}
\end{figure}

We have also calculated the optical absorption spectra of some ternary acetylides and
Cs$_{2}$Te (Figs. \ref{OptAbs-Pd} and \ref{OptAbs-Te}) 
in the Random Phase Approximation (RPA) using the YAMBO-code \cite{AMarini09}. 
All optical absorption
calculations have been performed with a resolution of ${\Delta}k<0.1$ {\AA}$^{-1}$,
and a Gaussian broadening of 0.03 Ry.
Note that for maximum absorption
the polarization of the light was parallel with rods in the ternary acetylides and parallel
with the crystallographic c-axis in Cs$_{2}$Te (see Fig. \ref{OptAbs-Te-XYZ}).
Due to the lack of norm-conserving pseudopotential for Pd, optical absorption spectra of
Na$_{2}$PdC$_{2}$ and Cs$_{2}$PdC$_{2}$ could not be calculated. In order to
associate these gaps with transition probabilities, a crude approximation
of these spectra using only planewaves with G=0 wave-vectors was attempted. 
It indicates absorption maxima at 1.8 and 2.6 eV for Na$_{2}$PdC$_{2}$ and 
Cs$_{2}$PdC$_{2}$, respectively (Fig. \ref{OptAbs-Pd}).
Unexpectedly, PBE0 \cite{CAdamo99} calculations at the same geometries result in about 1.0 eV
larger gaps than the experimental ones.
\begin{table}[h]
\caption{
Validation of the a, b and c lattice parameters on several test systems using the
PBE density functional, as described in the discussion. 
Orthorhombic and hexagonal
Cs$_{2}$C$_{2}$ are denoted as o-Cs$_{2}$C$_{2}$ and h-Cs$_{2}$C$_{2}$, respectively,
with structural parameters not very accurately determined due to the coexistence of the
two phases at any temperature.
}
\label{StructuralParams}
\begin{tabular}{ccccccc}
\hline
Compound,      & \multicolumn{6}{c}{Lattice Parameters ({\AA})}  \\
space-group \& & \multicolumn{3}{c}{EXPT} & \multicolumn{3}{c}{DFT} \\
reference      & a & b & c & a & b & c \\ 
\colrule
Cs (Im$\overline{3}$m)\cite{CSBarrett67}                 & 6.067 & 6.067 & 6.067 & 6.067 & 6.067  &  6.067  \\  
Te (P3$_{1}$21) \cite{NBouad03}                               & 4.526 & 4.526 & 5.920 & 4.458 & 4.458  &  5.925  \\  
Cs$_{2}$Te (Pnma)\cite{IScheweMiller02}                  & 9.512 & 5.838 &11.748 & 9.558 & 5.832  & 11.750  \\  
C (Fd$\overline{3}$m)\cite{MEStraumanis51}               & 3.567 & 3.567 & 3.567 & 3.573 & 3.573  & 3.573   \\
Na$_{2}$C$_{2}$ (I4$_{1}$/acd)\cite{SHemmersbach99}           & 6.778 & 6.778 &12.740 & 6.941 & 6.941  & 13.027  \\
o-Cs$_{2}$C$_{2}$ (Pnma)\cite{URuschewitz01}             & 9.545 & 5.001 &10.374 & 9.826 & 5.061  & 10.491  \\  
h-Cs$_{2}$C$_{2}$ (P$\overline{6}$2m)\cite{URuschewitz01}& 8.637 & 8.637 & 5.574 & 8.728 & 8.728  & 6.048   \\  
CsAgC$_{2}$(P4$_{2}$mmc)\cite{WKockelmann99}                  & 5.247 & 5.247 & 8.528 & 5.317 & 5.317  & 9.036   \\  
Na$_{2}$PdC$_{2}$ (P$\overline{3}$m1)\cite{HBilletter10} & 4.464 & 4.464 & 5.266 & 4.632 & 4.632  & 5.284   \\  
Cs$_{2}$PdC$_{2}$ (P$\overline{3}$m1)\cite{URuschewitz01A}& 5.624 & 5.624 & 5.298 & 5.804 & 5.804  & 5.265   \\  
Na$_{2}$TeC$_{2}$ (P$\overline{3}$m1)                    & -     & -     & -     & 4.767 & 4.767  & 6.102   \\  
Cs$_{2}$TeC$_{2}$ (P$\overline{3}$m1)                    & -     & -     & -     & 5.820 & 5.820  & 6.152   \\  
\hline
\end{tabular}
\end{table}
%
\begin{table}[h]
\caption{
Validation of C-C and M-C distances (M is transition-metal or metalloid element).
}
\label{Bondlengths}
\begin{tabular}{ccccc}
\hline
Compound,       & \multicolumn{2}{c}{d(C-C) ({\AA})} &  \multicolumn{2}{c}{d(M-C) ({\AA})} \\
Space-group \& ref. & EXPT & DFT  & EXPT & DFT \\
\hline
C (Fd$\overline{3}$m)\cite{MEStraumanis51}               & 1.544 & 1.547  & -     & -       \\
C$_{2}$H$_{2}$ (gas) \cite{JOverend60}                   & 1.203 & 1.203 & -     &  -      \\
Na$_{2}$C$_{2}$ (I41/acd) \cite{SHemmersbach99}          & 1.204 & 1.261 & -     &  -      \\
o-Cs$_{2}$C$_{2}$ (Pnma) \cite{URuschewitz01}            & 1.385 & 1.269  & -     & -       \\  
h-Cs$_{2}$C$_{2}$ (P$\overline{6}$2m)\cite{URuschewitz01}& 0.934 & 1.267  & -     & -       \\  
CsAgC$_{2}$ (P4$_{2}$mmc) \cite{WKockelmann99}           & 1.216 & 1.249 & 2.016 &  2.034  \\  
Na$_{2}$PdC$_{2}$ (P$\overline{3}$m1) \cite{HBilletter10}& 1.262 & 1.271  & 2.019 & 2.006    \\  
Cs$_{2}$PdC$_{2}$ (P$\overline{3}$m1) \cite{URuschewitz01A}& 1.260 & 1.280  & 2.019 & 1.993    \\  
Na$_{2}$TeC$_{2}$ (P$\overline{3}$m1)                    &  -    & 1.259  & -     & 2.422    \\  
Cs$_{2}$TeC$_{2}$ (P$\overline{3}$m1)                    & -     & 1.257  & -     & 2.452    \\  
\hline
\end{tabular}
\end{table}
%
%
\begin{table}[h]
\caption{
Experimental and calculated (DFT) properties of photoemissive surfaces
of validation materials:
workfunctions ($\Phi$), bandgaps at the $\Gamma$-point
E$_{g}(\Gamma)$ 
and surface energies ($\sigma$).
}
\label{ValidationWorkfunctions}
\begin{tabular}{ccccccc}
\hline  
Compound & \multicolumn{2}{c}{$\Phi$ (eV) } &\ & E$_{g}(\Gamma) (eV)$ & \ & $\sigma$ (eV/{\AA}$^{2}$) \\
surface  & EXPT & DFT & & DFT & & DFT \\
\hline  
Cs(100)         & 2.14 \cite{HBMichaelson77} &  2.00 &  & 0.29  &  & 0.005   \\
Te(001)         & 4.95 \cite{HBMichaelson77} &  5.02 &  & 0.54  &  & 0.036   \\
Cs$_{2}$Te(001) & 2.90-3.0 \cite{SLederer07} &  3.08 &  & 0.77  &  & 0.015   \\
Cs$_{2}$Te(010) & 2.90-3.0 \cite{SLederer07} &  2.90 &  & 1.04  &  & 0.014   \\
(Cs)Na$_{3}$KSb & 1.55 \cite{AHSommer68}     &  -    &  & -     &  & -  \\
K$_{2}$CsSb     & 1.9-2.1 \cite{IBazarov11,TVecchione11}  & - & & -   &   & -  \\
\hline  
\end{tabular}
\end{table}
%
\begin{table}[h]
\caption{
Calculated (DFT) properties of photoemissive surfaces 
of acetylide compounds:
workfunctions ($\Phi$), bandgaps at the $\Gamma$-point
E$_{g}(\Gamma)$ 
and surface energies ($\sigma$). 
Relaxed slabs refer to the relaxation of unrelaxed ones with the 
central 2 layers fixed. For h-Cs$_{2}$C$_{2}$(001)
and Na$_{2}$TeC$_{2}$(010), E$_{g}(\Gamma)$ $\approx0.05$ eV has been found for a 
single band above E$_{F}$ as well.
}
\label{TernaryAcetylidesWorkfunctions}
\begin{tabular}{ccccccc}
\hline  
Compound & \multicolumn{3}{c}{unrelaxed} & \multicolumn{3}{c}{relaxed} \\
and      & $\Phi$  & E$_{g}(\Gamma)$  & $\sigma$ & $\Phi$  & E$_{g}(\Gamma)$  & $\sigma$ \\
surface  & (eV) & (eV) & (eV/{\AA}$^{2}$) & (eV) & (eV) & (eV/{\AA}$^{2}$) \\
\hline  
o-Cs$_{2}$C$_{2}$(010)    & 2.80   & 1.25   & 0.023  &-     &-&-\\  
h-Cs$_{2}$C$_{2}$(001)    & 2.56   & 1.14   & 0.027  &-     &-&-\\  
Na$_{2}$PdC$_{2}$(001)    & 3.58   & 1.13   & 0.067  &-     &-&-\\  
Na$_{2}$PdC$_{2}$(110)    & 3.73   & 1.65   & 0.029  &4.17  & 2.34 & 0.024  \\  
Na$_{2}$PdC$_{2}$(010)    & 2.65   & 1.91   & 0.019  &2.68  & 2.45 & 0.017 \\  
Cs$_{2}$PdC$_{2}$(001)    & 2.90   & 1.43   & 0.046  &-     &-&-\\  
Cs$_{2}$PdC$_{2}$(110)    & 2.73   & 0.88   & 0.026  &2.73  & 1.16 & 0.022 \\  
Cs$_{2}$PdC$_{2}$(010)    & 1.33   & 0.78   & 0.015  &2.03  & 1.74 & 0.013 \\  
Na$_{2}$TeC$_{2}$(001)    & 3.40   & 1.03   & 0.029  &-     &-&-\\  
Na$_{2}$TeC$_{2}$(110)    & 3.80   & 0.91   & 0.025  &4.67  & 2.04 & 0.009 \\  
Na$_{2}$TeC$_{2}$(010)    & 2.75   & 1.43   & 0.015  &2.68  & 1.34 & 0.015 \\  
Cs$_{2}$TeC$_{2}$(001)    & 3.71   & 1.86   & 0.022  &-     &-&-\\  
Cs$_{2}$TeC$_{2}$(110)    & 2.77   & 0.77   & 0.020  &2.98  & 1.38 & 0.019 \\  
Cs$_{2}$TeC$_{2}$(010)    & 1.71   & 1.00   & 0.013  &2.44  & 1.63 & 0.009 \\  
\hline  
\end{tabular}
\end{table}
%
%
%
%
%
\section{Results and Discussion}
The optimization reveals that Cs$_{2}$TeC$_{2}$ has a very similar structure to other
compounds of the A$_{2}$MC$_{2}$ class.
Our DFT calculations predict that the electronic energy change in Eq. \ref{Cs2TeC2plusC2H2} is
${\Delta}$E = +1.1 eV per Cs$_{2}$TeC$_{2}$ unit, while that in the
alternative reaction of 
\begin{equation} \label{Cs2C2plusTe}
{\rm Cs}_{2}{\rm C}_{2}{\rm (cr)} + {\rm Te}{\rm (cr)} \rightarrow 
{\rm Cs}_{2}{\rm TeC}_{2}{\rm (cr)} 
\end{equation}
is ${\Delta}$E = -0.95 eV, indicating the stability of
the Cs$_{2}$TeC$_{2}$ crystal and an alternative synthesis route.
In fact the synthesis in Eq.
\ref{Cs2C2plusTe} is analogous to that of already existing A$_{2}$MC$_{2}$ compounds
\cite{HBilletter10}.
The predicted stability of a ternary acetylide with metalloid element (Te) instead of a 
transition metal for M in the A$_{2}$MC$_{2}$ formula is indicative of potential extension
of this class of materials with metalloids, while preserving the peculiar rod-like 
[MC$_{2}$]$_{\infty}$ substructures.
Our analysis at this point cannot exclude the existence of other structures for
Cs$_{2}$TeC$_{2}$. We have attempted to start the optimization of a 5 atomic unit
cell of Cs$_{2}$TeC$_{2}$ from several randomly chosen initial lattice parameters and
atomic positions. In all cases the formation of [TeC$_{2}$]$_{\infty}$ rods was evident after a
few hundred steps. Here we have relied on the fact that A$_{2}$MC$_{2}$ compounds
have been found only with hexagonal rod packing so far. Also, the structure of
h-Cs$_{2}$C$_{2}$ already contains the hexagonal rod-packing of the C$_{2}$
units leaving place for intercalatable atoms, such as Te, or transition metals,
between neighboring C$_{2}$-s along a rod.

One should also note that the linear chains of carbon atoms,
[C$_{2}$]$_{\infty}$ with alternating C-C and C$\equiv$C bonds (polycarbyne) or with uniform
C=C bonds (cumulenes, polyallenes), have long been a subject of theoretical and 
materials science interest \cite{SRoth04,FDiederich95}. 
However, unlike their hydrogenated analogue, [C$_{2}$H$_{2}$]$_{\infty}$
polyacetylene, containing alternating C-C and C=C bonds, 
famous for high electrical conductivity on the order of that of silver 
when doped \cite{CKChiang78,Nobel2000},
[C$_{2}$]$_{\infty}$ could not have been synthesized until a decade ago, proving the existence 
of [C$_{2}$]$_{n}$ with n$\in\approx$[200,300] \cite{FCataldo97}. Interestingly,
the efficient synthesis of [C$_{2}$]$_{n}$ involves copper-acetylides \cite{FCataldo97}.
Furthermore, copper-acetylides can also be used as the starting material for their synthesis
\cite{FCataldo99}, pointing to the intimate relationship of the rod-like
[MC$_{2}$]$_{\infty}$ substructures
in ternary-acetylides A$_{2}$MC$_{2}$ and AMC$_{2}$ to linear carbon chains.
Copper-acetylide molecules are also studied for their self assembly into extremely thin
nanowires \cite{KJudai06}. 
It is also important to note that while transition-metal
acetylides are known explosives, their alkalinated versions
AMC$_{2}$ and A$_{2}$MC$_{2}$ are not explosive at all and can survive heating up to
$\approx$ 500-600 $^{\rm o}$C and grinding \cite{HBilletter10,RBuschbeck11}.

\begin{figure*}[t!]
\resizebox*{6.8in}{!}{\includegraphics{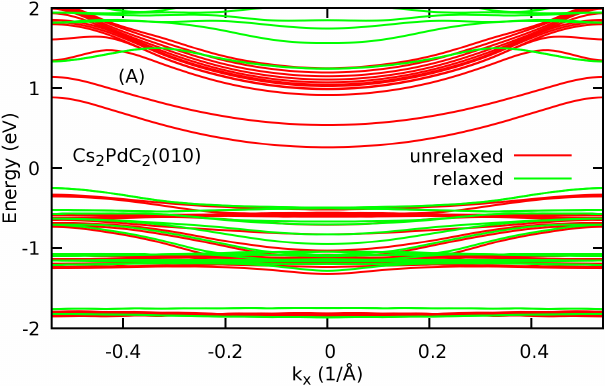}\includegraphics{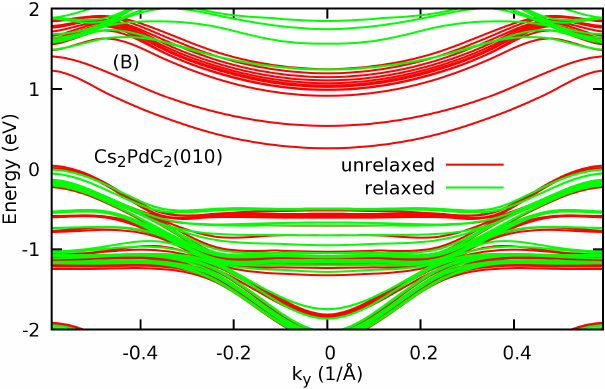}}
\resizebox*{6.8in}{!}{\includegraphics{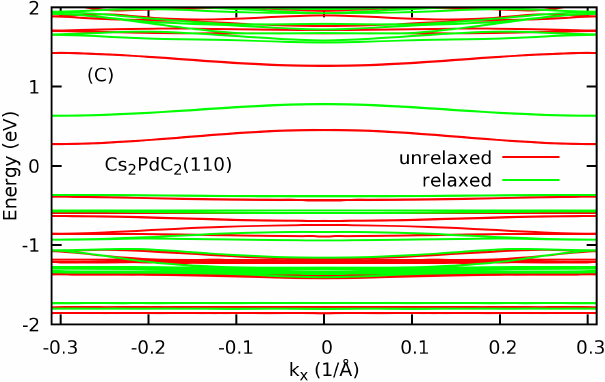}\includegraphics{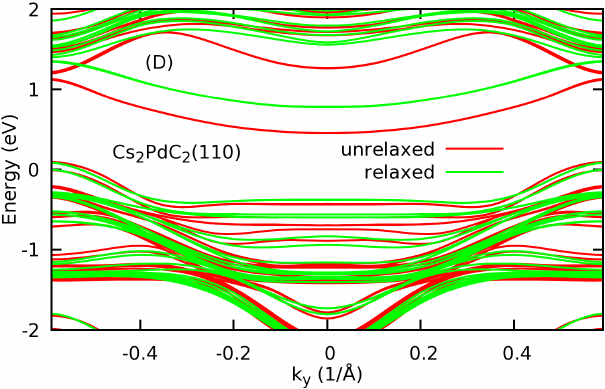}}
\resizebox*{6.8in}{!}{\includegraphics{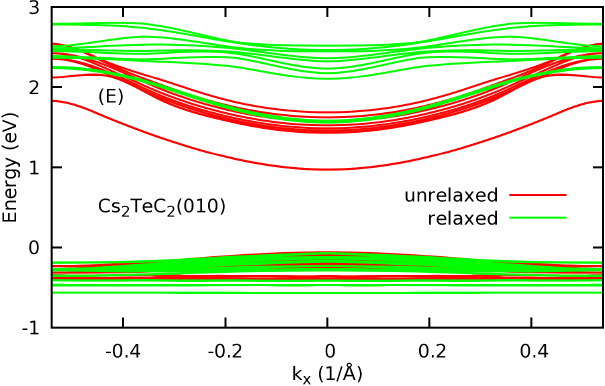}\includegraphics{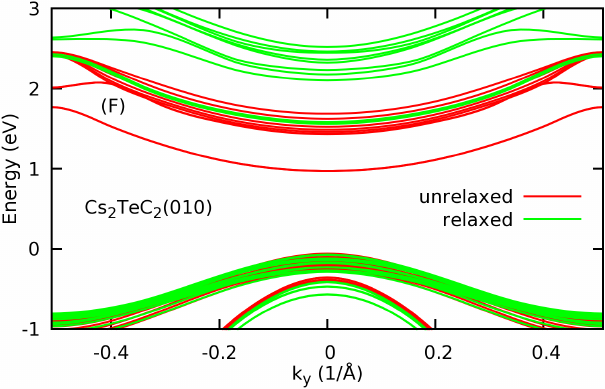}}
\resizebox*{6.8in}{!}{\includegraphics{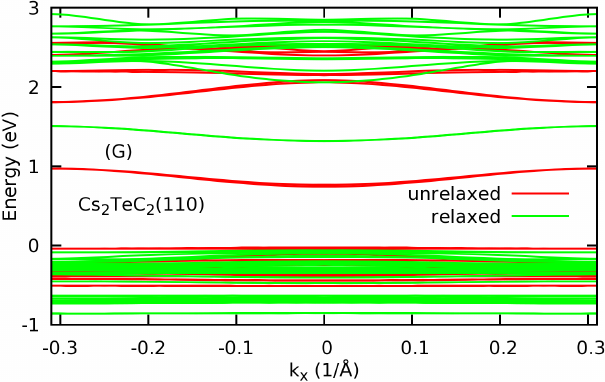}\includegraphics{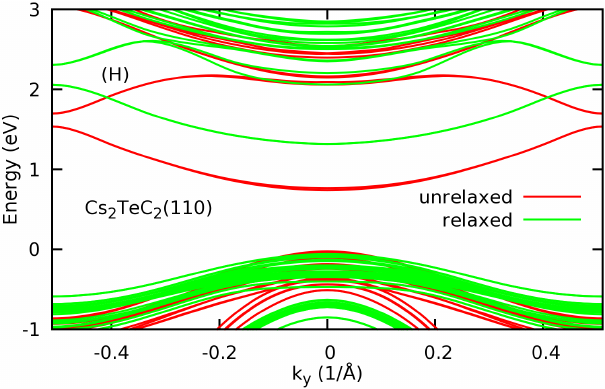}}
%
\caption{Electronic bands of the slabs of the (010) and (110) surfaces of 
Cs$_{2}$PdC$_{2}$ and Cs$_{2}$TeC$_{2}$ along the two orthogonal 
reciprocal surface lattice vectors k$_{x}$ (panels A, C, E and G) and k$_{y}$ 
(panels B, D, F and H). The [MC$_{2}$]$_{\infty}$ rods are parallel with the y direction
for all (010) and (110) surface slabs. Bands at both relaxed (green) and unrelaxed
(red) slabs are shown. The Fermi energy is at 0.
}
\label{SurfaceBands}
\end{figure*}

As it is indicated in Table \ref{TernaryAcetylidesWorkfunctions}, the workfunctions of different
surfaces of Cs$_{2}$TeC$_{2}$ have largely different values. 
Concerning the three most
important surfaces (Fig. \ref{Cs2TeC2surfaces}),
there is a $\approx$ 1 eV decrease as one goes from (001) through
(110) to (010) in each step, with workfunctions of 3.71, 2.77 and 1.71 eV, and 
surface energies of 0.022, 0.020 and 0.013 eV/{\AA}$^{2}$, respectively, for the 
unrelaxed surfaces. Relaxed surfaces have somewhat greater workfunction values, but 
still allowing for emission in the visible spectrum. 
Relaxation of the surface layers greatly influences the unoccupied bands, while the occupied
ones change significantly less, as indicated in Fig. \ref{SurfaceBands}.
Also note that the total energy
differences between relaxed and unrelaxed surfaces are small, for example they are 
only 0.3 eV for a whole Cs$_{2}$TeC$_{2}$(010) slab, i.e. about 0.01
eV/atom in the top surface layers which allows for thermal population 
of a great variety of surface structures at room temperature. 
In Cs$_{2}$TeC$_{2}$(010) and Na$_{2}$TeC$_{2}$(010)
surface relaxations may break the [TeC$_{2}$]$_{\infty}$ rods, while the rods stay intact
in Pd (or other transition metal) based ternary acetylides. In Cs$_{2}$PdC$_{2}$(010) and
Na$_{2}$PdC$_{2}$(010) the rods provide quasi rails along which Cs-s and Na-s 
can easily move due to thermal motion. This is also in accordance with the anomalous
broadening of peaks in the x-ray powder spectra of ternary acetylides \cite{URuschewitz06}.
Such an anomalous anisotropy of workfunction values is highly
unusual and represents a broad range of workfunction choice within a single material,
allowing for emission in ultraviolet, visible and near infrared radiation.
The lowest surface cleavage energy Cs$_{2}$TeC$_{2}$ surface, 
(010), has a similar surface energy as
those of Cs$_{2}$Te surfaces; it is, however, associated with a
much lower (by $\approx$ 1.3 eV) workfunction. 
The highly anisotropic properties of Cs$_{2}$TeC$_{2}$ are due to the relative orientation
of the rod-like [TeC$_{2}$]$_{\infty}$ substructures and the surfaces.
Surface energies
reveal that cutting the rods by cleaving the M-C bonds ((001) surface) 
is energetically disadvantageous, and it is
also disadvantageous to allow for rods to be directly exposed on the surface ((110) surface),
while cleavage between Cs atoms with rods embedded under the surface is the most
energetically favorable construct ((010) surface).
While numerous variants of surface coverages may exist at different temperatures 
that expose or cover rods by Cs on the 
surface, here we do not go beyond a single surface unit to study the energetics of
surfaces. The sticking of Cs to these surfaces may be a similarly important issue
here as in the case of cesiated III/V semiconductor surfaces (e.g. GaAs) \cite{MBesancon90}. 
As the rods are
twice negatively charged per MC$_{2}$ unit, we expect that the sticking of Cs
cations would be relatively strong.
%
%
%
%
\begin{figure}[t]
\resizebox*{3.4in}{!}{\includegraphics{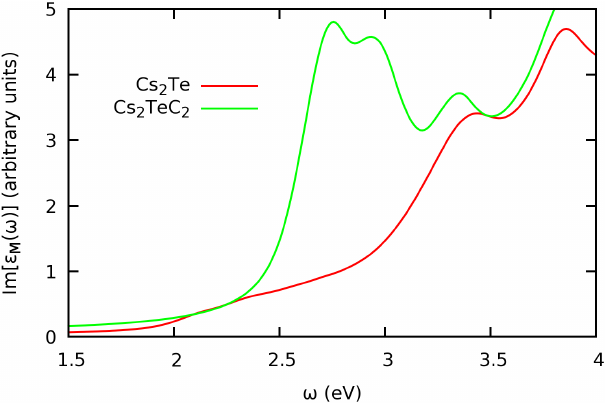}}
\caption{
Optical absorption spectra of bulk Cs$_{2}$Te and Cs$_{2}$TeC$_{2}$
from the imaginary part of the macroscopic dielectric function $\varepsilon_{M}(\omega)$.
The 4000 lowest energy planewaves were used to calculate intensities in the RPA
approximation. 
The polarization vector of the light is along the main
crystallographic axis (along the [TeC$_{2}$]$_{\infty}$ chains and the c-axis of
Cs$_{2}$Te).
}
\label{OptAbs-Te}
\end{figure}
%
%
\begin{figure}[t]
\resizebox*{3.4in}{!}{\includegraphics{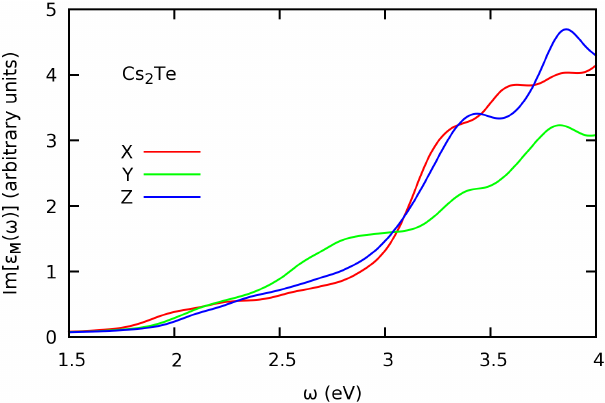}}
\resizebox*{3.4in}{!}{\includegraphics{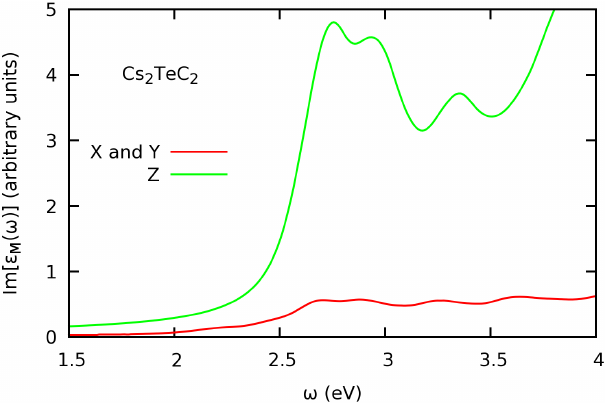}}
\caption{
Dependence of the optical absorption spectra of bulk Cs$_{2}$Te and Cs$_{2}$TeC$_{2}$
on the polarization of the incident light. The z-direction is along the main
crystallographic axis (c-axis), which is parallel with the [TeC$_{2}$]$_{\infty}$
chains in Cs$_{2}$TeC$_{2}$. While absorption in Cs$_{2}$TeC$_{2}$ is highly
anisotropic, with $\approx$ 9 times higher values for the z-direction than for the x and y
ones, there is no significant anisotropy
of absorption in Cs$_{2}$Te. Similar anisotropy can be seen in Na$_{2}$TeC$_{2}$ as
well, and likely in all ternary acetylides, due to the electric dipoles along the 
[MC$_{2}$]$_{\infty}$ chains.
}
\label{OptAbs-Te-XYZ}
\end{figure}
%
%
%

High anisotropy can be observed in Na$_{2}$PdC$_{2}$, Na$_{2}$TeC$_{2}$ and 
Cs$_{2}$PdC$_{2}$ as well, 
with somewhat smaller, 1.1-1.6 eV difference between the extremal surfaces. 
The type of the alkali atom very sensitively influences the workfunctions:
substituting Na with Cs results in more than 1 eV reduction of 
the workfunction on the (110) and (010) surfaces independently from the type
of the [MC$_{2}$]$_{\infty}$ chain, even though the M-C bonding in these chains
is very different.
One has to note that 
the Pd-C distance is significantly shorter than the Te-C in these compounds,
2.01 {\AA} vs. 2.45 {\AA}, respectively, while Te and Pd have very similar
covalent radii of $\approx 1.4$ {\AA} \cite{CBeatriz08}. The (001) surface energies
also indicate a much stronger Pd-C bond than Te-C one. 
While there is a $\sigma$-bond in both Pd-C and Te-C links between the 
2sp$^{1}$ hybrid orbital of the C atom 
and the 5sp$^{1}$ hybrids of Pd and Te (all oriented along the M-C-C line)
the Pd-C link is further strengthened
by strong back-donation of Pd 4d shell electrons to the antibonding 
$\pi$-orbitals of the C$_{2}^{2-}$ ions,
also associated with lengthening of the C-C bond \cite{HBilletter10}. 
Also note that Cs$_{2}$Te (and also Cs$_{2}$TeC$_{2}$) has the  
advantage over the formerly mentioned multi-alkali-antimonides that Cs is better bound in them 
allowing for longer operational lifetime \cite{DHDowell10}.
Another interesting comparison can be made to amorphous cesiated carbon films obtained
from the co-deposition of high-energy negatively charged carbon ions and Cs on silicon
substrates,
as the low, $\approx$ 1.1 eV workfunction in them might be associated with increased
acetylide ion concentration. However, there is no available data of how well Cs is bound in
these systems \cite{YWKo97}.

In order to estimate the quantum-yield of Cs$_{2}$TeC$_{2}$ relative to Cs$_{2}$Te,
we have calculated their optical absorption spectra (Fig. \ref{OptAbs-Te}) 
using the lowest energy 4000 planewaves at which the spectrum becomes saturated against
further increase of the number of planewaves.
The spectra indicate that acetylation of Cs$_{2}$Te shifts its first absorption peak in the
visible region to 2.7 eV, while preserving the same absorption intensity.
This comparison suggests that Cs$_{2}$TeC$_{2}$ may have
similarly high quantum efficiency as that of Cs$_{2}$Te, however, even for visible and
potentially also for near infrared photons. The bandgaps at the $\Gamma$ point
of Cs$_{2}$TeC$_{2}$ surfaces (see Table \ref{TernaryAcetylidesWorkfunctions}) 
also support that photon-energies near the work-function are
sufficient to induce emission in this material. 
An interesting characteristics of ternary acetylides is the extensive presence of flat bands (see 
Figs. \ref{bands-Pd} and \ref{bands-Te}). While there are some flat band parts in Cs$_{2}$Te as
well, such a feauture is much more characteristic for ternary acetylides. Flat bands greatly
increase the density of states for some spectral regions thus they contribute to increased
absorption of light.
Interestingly, not only the workfunctions of these materials show high anisotropy, but also
their optical absorption (see Fig. \ref{OptAbs-Te-XYZ}). The optical absorption is almost a
magnitude greater when the light's polarization vector is parallel with the
[MC$_{2}$]$_{\infty}$ rods.
This property can allow for example for the generation of pulsed electron beams when these
surfaces are illuminated by circularly polarized light. Several other optical applications
can be envisioned based on the anisotropy of optical absorption in ternary acetylides, such
as polar-filters and optical switching elements.

It is also important to call attention
to the rest of the ternary acetylides as valuable photoemissive materials. 
For example the already synthesized Cs$_{2}$PdC$_{2}$(010) \cite{URuschewitz01A} material
is predicted here to have a very low 1.33-2.03 eV workfunction even smaller than     
that of Cs$_{2}$TeC$_{2}$(010) and a similar density of states.

While it may be difficult to lower the workfunction into the infrared spectral domain (below
1.5 eV), multiphoton absorption of infrared light may still provide a way to photo-emission
in this domain as well. Strong multiphoton absorption of organic and inorganic compounds
with acetylide units is well known \cite{GSHe08}, for example in platinum
acetylides \cite{KSonogashira78}, the analogy suggests that multiphoton absorption may be
strong in ternary acetylides as well. Multiphoton absorption
happens via simultaneous absorption of multiple photons without the need of real
intermediate states as opposed to cascaded multiple step one-photon absorptions
\cite{GSHe08}. These
latter ones are also possible in ternary acetylides as there are surface states 
1-1.5 eV above the Fermi level as indicated in Fig. \ref{SurfaceBands}.

Emission from A$_{2}$MC$_{2}$(001) surfaces (rods perpendicular to surface) 
may especially be suitable for generating 
low transverse emittance electron beams \cite{KNemeth10} as excited electrons are expected
to be guided along the [MC$_{2}$]$_{\infty}$ rods while traveling from inside the bulk 
of the cathode towards the surface whereby not being scattered side-wise, analogously to
needle-array cathodes of field emission \cite{RGanter08}.

\section{Conclusions}
In the present work we have demonstrated unique photoemissive properties of ternary acetylides,
such as low workfunctions, high workfunction and optical absorption 
anisotropy and high quantum yield. 
We have also demonstrated how the acetylation of the seasoned Cs$_{2}$Te photocathode
material leads to significantly lowered workfunction while preserving its high quantum yield.
 
\section{Acknowledgements}
The authors gratefully acknowledge A. Zholents and K. Attenkofer (APS/Argonne)
for helpful discussions and thank NERSC (U.S. DOE DE-AC02-05CH11231) for
the use of computational resources. 
J. Z. Terdik thanks A. Zholents for support.
This research was supported
by the U.S. DOE Office of Science, under contract No.
DE-AC02-06CH11357, and also by the National Science Foundation (No. PHY-0969989).

\noindent
{$\ast$ Nemeth@ANL.Gov}
%

\begin{thebibliography}{40}
\expandafter\ifx\csname natexlab\endcsname\relax\def\natexlab#1{#1}\fi
\expandafter\ifx\csname bibnamefont\endcsname\relax
  \def\bibnamefont#1{#1}\fi
\expandafter\ifx\csname bibfnamefont\endcsname\relax
  \def\bibfnamefont#1{#1}\fi
\expandafter\ifx\csname citenamefont\endcsname\relax
  \def\citenamefont#1{#1}\fi
\expandafter\ifx\csname url\endcsname\relax
  \def\url#1{\texttt{#1}}\fi
\expandafter\ifx\csname urlprefix\endcsname\relax\def\urlprefix{URL }\fi
\providecommand{\bibinfo}[2]{#2}
\providecommand{\eprint}[2][]{\url{#2}}

\bibitem[{\citenamefont{{D. H. Dowell {\it et.al}}}(2010)}]{DHDowell10}
\bibinfo{author}{\bibnamefont{{D. H. Dowell {\it et.al}}}},
  \bibinfo{journal}{Nuclear Instruments and Methods in Physics Research A}
  \textbf{\bibinfo{volume}{622}}, \bibinfo{pages}{685} (\bibinfo{year}{2010}).

\bibitem[{\citenamefont{Taft and Apker}(1953)}]{ETaft53}
\bibinfo{author}{\bibfnamefont{E.}~\bibnamefont{Taft}} \bibnamefont{and}
  \bibinfo{author}{\bibfnamefont{L.}~\bibnamefont{Apker}}, \bibinfo{journal}{J.
  Opt. Soc. Am.} \textbf{\bibinfo{volume}{43}}, \bibinfo{pages}{81}
  (\bibinfo{year}{1953}).

\bibitem[{\citenamefont{{A. di Bona {\it et.al}}}(1996)}]{AdiBona96}
\bibinfo{author}{\bibnamefont{{A. di Bona {\it et.al}}}}, \bibinfo{journal}{J.
  Appl. Phys.} \textbf{\bibinfo{volume}{80}}, \bibinfo{pages}{3024}
  (\bibinfo{year}{1996}).

\bibitem[{\citenamefont{Ruschewitz}(2001)}]{URuschewitz01A}
\bibinfo{author}{\bibfnamefont{U.}~\bibnamefont{Ruschewitz}},
  \bibinfo{journal}{Z. Anorg. Allg. Chem.} \textbf{\bibinfo{volume}{627}},
  \bibinfo{pages}{1231} (\bibinfo{year}{2001}).

\bibitem[{\citenamefont{{H. Billetter {\it et.al}}}(2010)}]{HBilletter10}
\bibinfo{author}{\bibnamefont{{H. Billetter {\it et.al}}}},
  \bibinfo{journal}{Z. Anorg. Allg. Chem.} \textbf{\bibinfo{volume}{636}},
  \bibinfo{pages}{1834} (\bibinfo{year}{2010}).

\bibitem[{\citenamefont{{W. Kockelmann and U.
  Ruschewitz}}(1999)}]{WKockelmann99}
\bibinfo{author}{\bibnamefont{{W. Kockelmann and U. Ruschewitz}}},
  \bibinfo{journal}{Angew. Chem. Int. ed.} \textbf{\bibinfo{volume}{38}},
  \bibinfo{pages}{3495} (\bibinfo{year}{1999}).

\bibitem[{\citenamefont{{U. Ruschewitz }}(2006)}]{URuschewitz06}
\bibinfo{author}{\bibnamefont{{U. Ruschewitz }}}, \bibinfo{journal}{Z. Anorg.
  Allg. Chem.} \textbf{\bibinfo{volume}{632}}, \bibinfo{pages}{705}
  (\bibinfo{year}{2006}).

\bibitem[{\citenamefont{{P. Gianozzi {\it et.al}}}(2009)}]{QE}
\bibinfo{author}{\bibnamefont{{P. Gianozzi {\it et.al}}}}, \bibinfo{journal}{J.
  Phys.: Condens. Matter} \textbf{\bibinfo{volume}{21}},
  \bibinfo{pages}{395502} (\bibinfo{year}{2009}),
  \bibinfo{note}{http://www.quantum-espresso.org}.

\bibitem[{\citenamefont{{J. P. Perdew {\it et.al}}}(1996)}]{PBE}
\bibinfo{author}{\bibnamefont{{J. P. Perdew {\it et.al}}}},
  \bibinfo{journal}{Phys. Rev. Lett.} \textbf{\bibinfo{volume}{77}},
  \bibinfo{pages}{3865} (\bibinfo{year}{1996}).

\bibitem[{\citenamefont{{C. J. Fall {\it et.al}}}(1999)}]{CJFall99}
\bibinfo{author}{\bibnamefont{{C. J. Fall {\it et.al}}}}, \bibinfo{journal}{J.
  Phys.: Cond. Mat.} \textbf{\bibinfo{volume}{11}}, \bibinfo{pages}{2689}
  (\bibinfo{year}{1999}).

\bibitem[{\citenamefont{{S. Hemmersbach {\it et.al}}}(2001)}]{SHemmersbach01}
\bibinfo{author}{\bibnamefont{{S. Hemmersbach {\it et.al}}}},
  \bibinfo{journal}{Chem. Eur. J.} \textbf{\bibinfo{volume}{7}},
  \bibinfo{pages}{1952} (\bibinfo{year}{2001}).

\bibitem[{\citenamefont{{A. Marini {\it et.al}}}(2009)}]{AMarini09}
\bibinfo{author}{\bibnamefont{{A. Marini {\it et.al}}}},
  \bibinfo{journal}{Comp. Phys. Comm.} \textbf{\bibinfo{volume}{180}},
  \bibinfo{pages}{1392} (\bibinfo{year}{2009}),
  \bibinfo{note}{http://www.yambo-code.org}.

\bibitem[{\citenamefont{{C. Adamo {\it et.al}}}(1999)}]{CAdamo99}
\bibinfo{author}{\bibnamefont{{C. Adamo {\it et.al}}}}, \bibinfo{journal}{J.
  Chem. Phys.} \textbf{\bibinfo{volume}{110}}, \bibinfo{pages}{6158}
  (\bibinfo{year}{1999}).

\bibitem[{\citenamefont{{C. S. Barrett}}(1956)}]{CSBarrett67}
\bibinfo{author}{\bibnamefont{{C. S. Barrett}}}, \bibinfo{journal}{Acta Cryst.}
  \textbf{\bibinfo{volume}{9}}, \bibinfo{pages}{671} (\bibinfo{year}{1956}).

\bibitem[{\citenamefont{{N. Bouad {\it et.al}}}(2003)}]{NBouad03}
\bibinfo{author}{\bibnamefont{{N. Bouad {\it et.al}}}}, \bibinfo{journal}{J.
  Solid State Chem.} \textbf{\bibinfo{volume}{173}}, \bibinfo{pages}{189}
  (\bibinfo{year}{2003}).

\bibitem[{\citenamefont{{I. Schewe-Miller {\it
  et.al}}}(2002)}]{IScheweMiller02}
\bibinfo{author}{\bibnamefont{{I. Schewe-Miller {\it et.al}}}},
  \bibinfo{journal}{Golden Book of Phase Transitions 5, Wroclaw}
  \textbf{\bibinfo{volume}{1}}, \bibinfo{pages}{123} (\bibinfo{year}{2002}).

\bibitem[{\citenamefont{{M. E. Straumanis {\it et.al}}}(1951)}]{MEStraumanis51}
\bibinfo{author}{\bibnamefont{{M. E. Straumanis {\it et.al}}}},
  \bibinfo{journal}{J. Am. Chem. Soc.} \textbf{\bibinfo{volume}{73}},
  \bibinfo{pages}{5643} (\bibinfo{year}{1951}).

\bibitem[{\citenamefont{{S. Hemmersbach {\it et.al}}}(1999)}]{SHemmersbach99}
\bibinfo{author}{\bibnamefont{{S. Hemmersbach {\it et.al}}}},
  \bibinfo{journal}{Z. Anorg. Allg. Chem.} \textbf{\bibinfo{volume}{625}},
  \bibinfo{pages}{1440} (\bibinfo{year}{1999}).

\bibitem[{\citenamefont{{U. Ruschewitz {\it et.al}}}(2001)}]{URuschewitz01}
\bibinfo{author}{\bibnamefont{{U. Ruschewitz {\it et.al}}}},
  \bibinfo{journal}{Z. Anorg. Allg. Chem.} \textbf{\bibinfo{volume}{627}},
  \bibinfo{pages}{513} (\bibinfo{year}{2001}).

\bibitem[{\citenamefont{{John Overend }}(1960)}]{JOverend60}
\bibinfo{author}{\bibnamefont{{John Overend }}}, \bibinfo{journal}{Trans.
  Faraday Soc.} \textbf{\bibinfo{volume}{56}}, \bibinfo{pages}{310}
  (\bibinfo{year}{1960}).

\bibitem[{\citenamefont{{H. B. Michaelson}}(1977)}]{HBMichaelson77}
\bibinfo{author}{\bibnamefont{{H. B. Michaelson}}}, \bibinfo{journal}{J. Appl.
  Phys.} \textbf{\bibinfo{volume}{48}}, \bibinfo{pages}{4729}
  (\bibinfo{year}{1977}).

\bibitem[{\citenamefont{{S. Lederer {\it et.al}}}(2007)}]{SLederer07}
\bibinfo{author}{\bibnamefont{{S. Lederer {\it et.al}}}}, in
  \emph{\bibinfo{booktitle}{Proc. FEL 2007}} (\bibinfo{year}{2007}), p.
  \bibinfo{pages}{457}.

\bibitem[{\citenamefont{Sommer}(1968)}]{AHSommer68}
\bibinfo{author}{\bibfnamefont{A.}~\bibnamefont{Sommer}},
  \emph{\bibinfo{title}{Photoemissive Materials}} (\bibinfo{publisher}{John
  Wiley \& Sons}, \bibinfo{address}{New York}, \bibinfo{year}{1968}).

\bibitem[{\citenamefont{{I. Bazarov {\it et.al}}}(2011)}]{IBazarov11}
\bibinfo{author}{\bibnamefont{{I. Bazarov {\it et.al}}}},
  \bibinfo{journal}{Appl. Phys. Lett.} \textbf{\bibinfo{volume}{98}},
  \bibinfo{pages}{224101} (\bibinfo{year}{2011}).

\bibitem[{\citenamefont{{T. Vecchione {\it et.al}}}(2011)}]{TVecchione11}
\bibinfo{author}{\bibnamefont{{T. Vecchione {\it et.al}}}},
  \bibinfo{journal}{Appl. Phys. Lett.} \textbf{\bibinfo{volume}{99}},
  \bibinfo{pages}{034103} (\bibinfo{year}{2011}).

\bibitem[{\citenamefont{Roth and Carroll}(2004)}]{SRoth04}
\bibinfo{author}{\bibfnamefont{S.}~\bibnamefont{Roth}} \bibnamefont{and}
  \bibinfo{author}{\bibfnamefont{D.}~\bibnamefont{Carroll}},
  \emph{\bibinfo{title}{One Dimensional Metals: conjugated polymers, organic
  crystals, carbon nanotubes}} (\bibinfo{publisher}{Wiley-VCH},
  \bibinfo{address}{Weinheim}, \bibinfo{year}{2004}).

\bibitem[{\citenamefont{Diederich}(1995)}]{FDiederich95}
\bibinfo{author}{\bibfnamefont{F.}~\bibnamefont{Diederich}}, in
  \emph{\bibinfo{booktitle}{Modern Acetylene Chemistry}}, edited by
  \bibinfo{editor}{\bibfnamefont{P.~J.} \bibnamefont{Stang}} \bibnamefont{and}
  \bibinfo{editor}{\bibfnamefont{F.}~\bibnamefont{Diederich}}
  (\bibinfo{publisher}{VCH}, \bibinfo{address}{Weinheim},
  \bibinfo{year}{1995}).

\bibitem[{\citenamefont{{C. K. Chiang {\it et.al}}}(1978)}]{CKChiang78}
\bibinfo{author}{\bibnamefont{{C. K. Chiang {\it et.al}}}},
  \bibinfo{journal}{J. Am. Chem. Soc.} \textbf{\bibinfo{volume}{100}},
  \bibinfo{pages}{1013} (\bibinfo{year}{1978}).

\bibitem[{Nob()}]{Nobel2000}
\bibinfo{note}{{Nobel Prize in Chemistry, 2000, to A. J. Heeger, A. G.
  MacDiarmid and H. Shirakawa}}.

\bibitem[{\citenamefont{Cataldo}(1997)}]{FCataldo97}
\bibinfo{author}{\bibfnamefont{F.}~\bibnamefont{Cataldo}},
  \bibinfo{journal}{Polymer International} \textbf{\bibinfo{volume}{44}},
  \bibinfo{pages}{191} (\bibinfo{year}{1997}).

\bibitem[{\citenamefont{Cataldo}(1999)}]{FCataldo99}
\bibinfo{author}{\bibfnamefont{F.}~\bibnamefont{Cataldo}},
  \bibinfo{journal}{Polymer International} \textbf{\bibinfo{volume}{48}},
  \bibinfo{pages}{15} (\bibinfo{year}{1999}).

\bibitem[{\citenamefont{{K. Judai {\it et.al}}}(2006)}]{KJudai06}
\bibinfo{author}{\bibnamefont{{K. Judai {\it et.al}}}}, \bibinfo{journal}{Adv.
  Mater.} \textbf{\bibinfo{volume}{18}}, \bibinfo{pages}{2842}
  (\bibinfo{year}{2006}).

\bibitem[{\citenamefont{{R. Buschbeck {\it et.al}}}(2011)}]{RBuschbeck11}
\bibinfo{author}{\bibnamefont{{R. Buschbeck {\it et.al}}}},
  \bibinfo{journal}{Coordination Chemistry Reviews}
  \textbf{\bibinfo{volume}{255}}, \bibinfo{pages}{241} (\bibinfo{year}{2011}).

\bibitem[{\citenamefont{{M. Besan{\c{c}}on {\it et.al}}}(1990)}]{MBesancon90}
\bibinfo{author}{\bibnamefont{{M. Besan{\c{c}}on {\it et.al}}}},
  \bibinfo{journal}{Surface Science} \textbf{\bibinfo{volume}{236}},
  \bibinfo{pages}{23} (\bibinfo{year}{1990}).

\bibitem[{\citenamefont{{C. Beatriz {\it et.al}}}(2008)}]{CBeatriz08}
\bibinfo{author}{\bibnamefont{{C. Beatriz {\it et.al}}}},
  \bibinfo{journal}{Dalton Trans.} \textbf{\bibinfo{volume}{21}},
  \bibinfo{pages}{2832} (\bibinfo{year}{2008}).

\bibitem[{\citenamefont{{Y. W. Ko and S. I. Kim}}(1997)}]{YWKo97}
\bibinfo{author}{\bibnamefont{{Y. W. Ko and S. I. Kim}}}, \bibinfo{journal}{J.
  Appl. Phys.} \textbf{\bibinfo{volume}{82}}, \bibinfo{pages}{2631}
  (\bibinfo{year}{1997}).

\bibitem[{\citenamefont{{Guang S. He {\it et.al}}}(2008)}]{GSHe08}
\bibinfo{author}{\bibnamefont{{Guang S. He {\it et.al}}}},
  \bibinfo{journal}{Chem. Rev.} \textbf{\bibinfo{volume}{108}},
  \bibinfo{pages}{1245} (\bibinfo{year}{2008}).

\bibitem[{\citenamefont{{K. Sonogashira {\it et.al}}}(1978)}]{KSonogashira78}
\bibinfo{author}{\bibnamefont{{K. Sonogashira {\it et.al}}}},
  \bibinfo{journal}{J. Organomet. Chem.} \textbf{\bibinfo{volume}{145}},
  \bibinfo{pages}{101} (\bibinfo{year}{1978}).

\bibitem[{\citenamefont{{K. N{\'e}meth {\it et.al}}}(2010)}]{KNemeth10}
\bibinfo{author}{\bibnamefont{{K. N{\'e}meth {\it et.al}}}},
  \bibinfo{journal}{Phys. Rev. Lett.} \textbf{\bibinfo{volume}{104}},
  \bibinfo{pages}{046801} (\bibinfo{year}{2010}).

\bibitem[{\citenamefont{{R. Ganter {\it et.al}}}(2008)}]{RGanter08}
\bibinfo{author}{\bibnamefont{{R. Ganter {\it et.al}}}},
  \bibinfo{journal}{Phys.\ Rev.\ Lett.} \textbf{\bibinfo{volume}{100}},
  \bibinfo{pages}{064801} (\bibinfo{year}{2008}).

\end{thebibliography}
%

\end{document}